\documentclass[aps,prl,reprint,superscriptaddress]{revtex4-2}

\usepackage{todonotes}
\usepackage[mathscr]{eucal}
\usepackage{blindtext}
\usepackage{amsmath}
\usepackage{upgreek}
\usepackage{hyperref}
\newcommand{\ie}{\emph{i.e.}}
\newcommand{\eg}{\emph{e.g.}}

\newcommand{\kBT}{{k_\te B T}}
\newcommand{\kB}{{k_\te B}}

\newcommand{\te}[1]{\mathrm{#1}}
\newcommand{\df}{\mathrm{d}}
\newcommand{\vv}[1]{\boldsymbol{#1}}
\newcommand{\ev}[1]{\langle #1 \rangle}

\newcommand{\pt}[1]{\left( #1 \right)}

\begin{document}


\title{Probing scale-dependent liveliness with nonequilibrium thermospectroscopy}



\author{Joscha Mecke}
\email[]{joscha.mecke@szu.edu.cn}
\affiliation{ 
	Institute for Advanced Study, Shenzhen University, 518060 Shenzhen, P.R.~China
}%
\affiliation{ 
	College of Physics and Optoelectronic Engineering, Shenzhen University, 518060 Shenzhen, P.R.~China
}
\author{Klaus Kroy}
\email[]{klaus.kroy@itp.uni-leipzig.de}
\affiliation{%
	Institute for Theoretical Physics, University of Leipzig, Postfach 100 920, D-04009 Leipzig, Germany}


\date{\today}

\begin{abstract}
	Probing the spatially heterogeneous activity across scales is a major challenge in living matter. %
	Energy injection at diverse length scales leads to mode coupling, inter-modal energy transfer, and entropy production. %
	We demonstrate the emergence of multiple effective (``active'') temperatures in nonequilibrium molecular- and Brownian-dynamics simulations of an active polymer. %
	Via a generalised Langevin equation for a labelled monomer we identify spectral noise temperatures and their relation to the underlying activity landscape. %
	A harmonic trap of variable stiffness can serve as a minimally invasive prototypical spectroscopic device to selectively scan through the emergent effective temperatures and thereby resolve the scale-dependent activity. %
\end{abstract}


\maketitle

Active matter consumes energy on a particle scale in order to generate nonequilibrium mesoscale dynamics.  Natural examples include bacteria, sperm, or algae~\cite{elgeti2015physics, gompper20202020}. %
During intra and extra-cellular transport processes, molecular motors harvest chemical energy in order to continuously apply forces to the cytoskeletal network leading to significant shape fluctuations~\cite{brangwynne2008nonequilibrium, sciortino2025active, koehler2011structure, ito2026force}, and, possibly, the movement of muscles more than $10^6$ times larger then the energy injecting unit~\cite{fang2019nonequilibrium}. %
Observing active or driven dynamics on mesoscopic or macroscopic scales generally fails to provide insights into the microscopic working mechanism. In particular, the underlying driving protocols such as inhomogeneous motor or fuel distributions from, \eg, membrane fluctuations, remain elusive. %
How energy gets transported and manifests an organism's liveliness across scales, or how local activity shapes large-scale dynamics, has only been understood for very specific models. A universal method to probe and meaningfully characterise multiscale activity remains lacking.

\begin{figure}[b]
	\centering
	\includegraphics[width=\columnwidth]{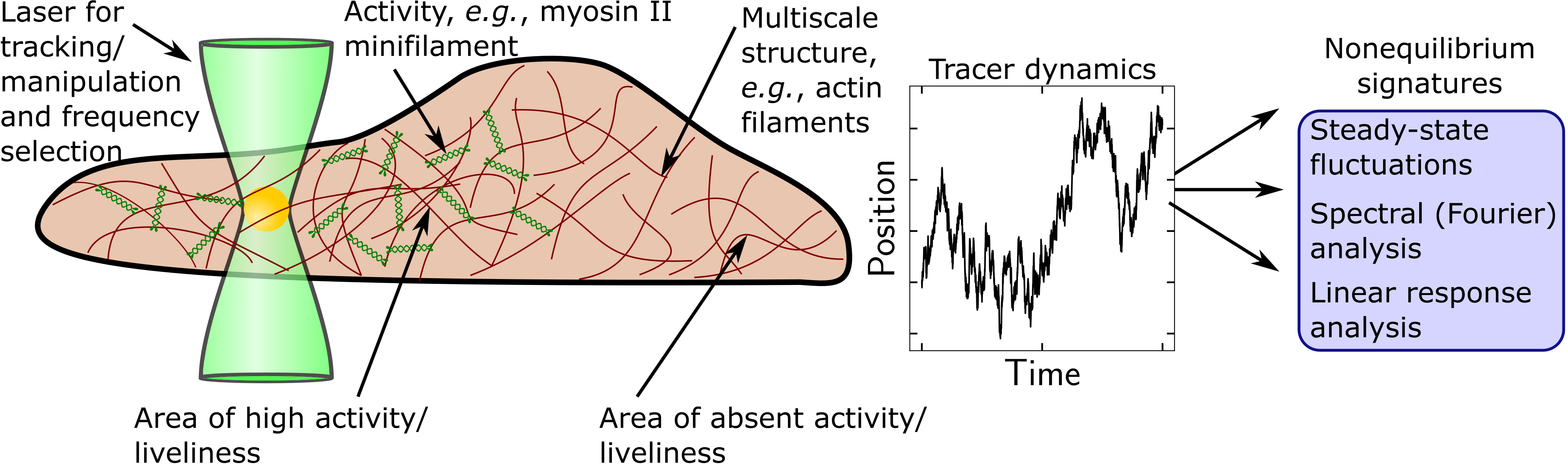}
	\caption{%
		Probing nonequilibrium tracer fluctuations in living cells. %
		The intra cellular network experiences active driving at various scales. %
		The fluctuating tracer trajectory encodes signatures of the corresponding activity landscape comprising areas of high activity (liveliness) and areas without local active contributions (dead areas). %
	}
	\label{fig:sketch}
\end{figure}

In driven and active matter, the spontaneous thermalisation of all interacting degrees of freedom at a common temperature, acknowledged by the zeroth law of thermodynamics, is permanently prohibited. %
However, a useful set of effective temperatures can often still be defined, via the conventional route, using the second law of thermodynamics, if the observed fluctuations could more or less be mimicked by the action of some virtual equilibrium reservoirs~\cite{wiese2024modeling}. %
While this approach relies on approximately Gaussian statistics of the fluctuations of the considered degrees of freedom, equipartition is not necessary. Correlations due to the energy propagation across scales~\cite{battle2016broken, bacanu2023inferring, gladrow2016broken} notwithstanding, selected wavelengths may sometimes carry much more energy than others~\cite{neu2024irreversible}, suggesting the usefulness of corresponding effective temperatures, which can vary with wavenumber or frequency~\cite{falasco2014effective, massana2024multiple}. %
Systematically, such mesoscale effective temperatures can be obtained via non-isothermal coarse-graining, starting from a heterogeneous molecular temperature or local activity field~\cite{falasco2016nonisothermal}. %
In principle, effective temperatures of mesoscale degrees of freedom can thus be employed to probe their possibly non-local and multiscale driving. %
But despite its broad spectrum of potential applications (Fig.~\ref{fig:sketch}), a direct application of this concept has so far largely remained an open challenge.

In this Letter, we explicitly develop such thermospectroscopic method~\cite{geiss2020thermometry, falasco2014non}, addressing a key inverse problem in active matter. Namely, inferring salient features of a multiscale driven or active environment from effective temperatures of tracer particles. %
This allows us not only to identify if a system is out of equilibrium, but moreover also to reconstruct hidden activity landscapes and distinguish ``lively'' from ``dead'' regions or scales, using only the fluctuating trajectory of a tracer particle. %
We investigate how energy is distributed, transported, and dissipated across spatial scales, as encoded in the entropy production rates and various effective temperatures, suggesting a range of minimally invasive applications in synthetic and biological matter far from equilibrium. %
We develop the concept to a proof-of-principle level by comparing a fully analytical generalised Langevin formalism to molecular and Brownian dynamics simulations.
The framework can straight forwardly be extended to any interacting many-body system out of equilibrium, the configurational dynamics of which can be decomposed into Fourier modes and thus provides a key step forward in solving the inverse problem and advancing the understanding of multiscale activity. %

\begin{figure}[t]
	\centering
	\includegraphics[width=\columnwidth]{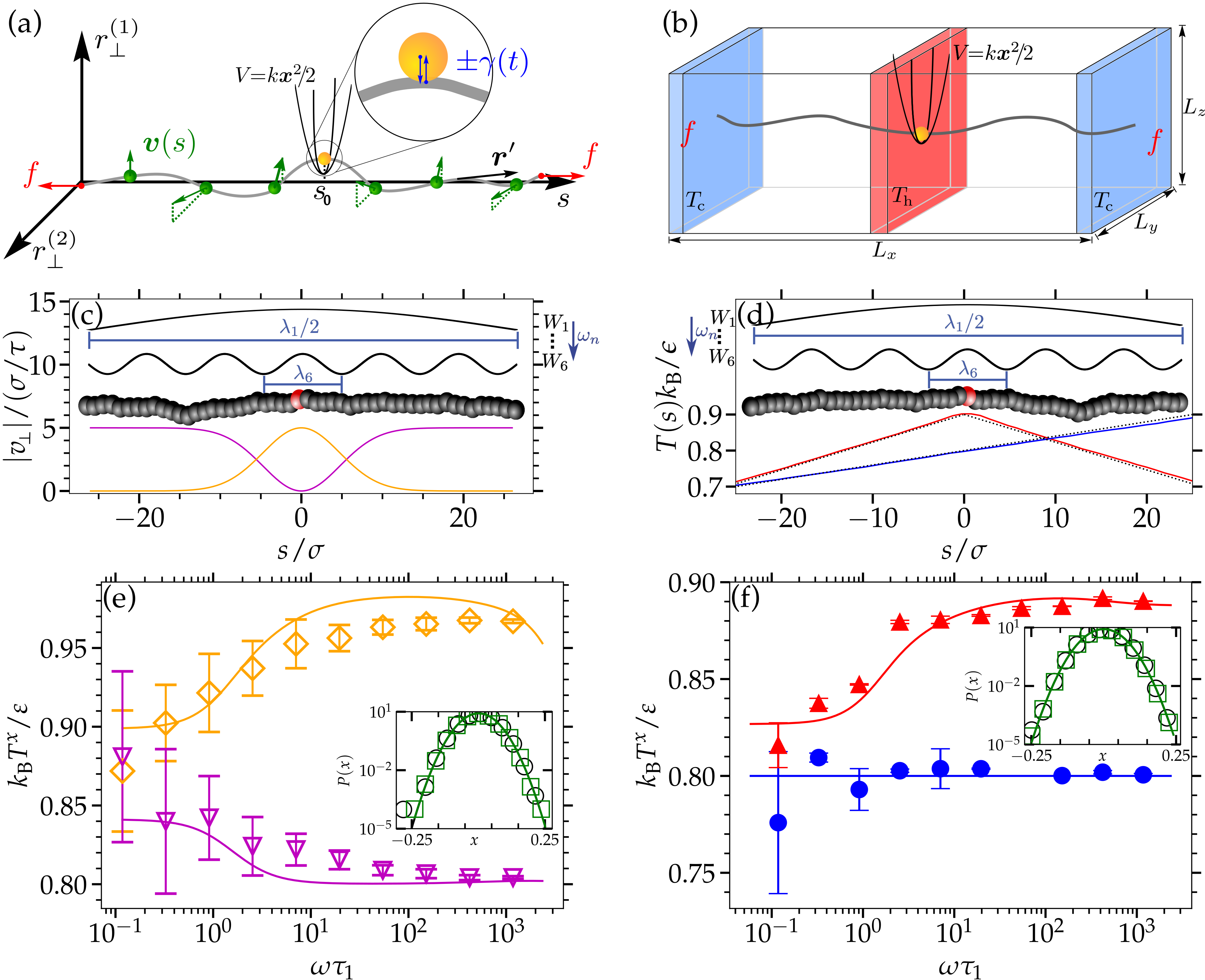}
	\caption{%
		Emergence of multiple effective temperatures in active Brownian~(a) and nonequilibrium molecular-dynamics simulations in a non-isothermal Lennard-Jones solvent~(b). %
		(c)~Swim speeds at different positions along the polymer chain. %
		Insets: Sketch featuring the first and sixth bending modes of the polymer, showing that at high frequencies, the main modal contribution stems from the direct vicinity of the tracer. %
		(d)~Triangular (red) and linear (blue) local molecular temperature profile in molecular dynamics simulations. Dotted lines show predictions by the stationary heat equation. %
		Brownian~(e) and molecular~(f) dynamics simulations versus our prediction from Eq.~\eqref{eq:effective_noise_temperature_dissipation}. Insets confirm a Gaussian probability distribution of the tracer positions. Colour as in (c)-(d). %
	}
	\label{fig:simulation}
\end{figure}

\emph{Emergence of multiple temperatures in a multiscale nonequilibrium environment.}\textemdash %
As a paradigm, we study the positional fluctuations of a harmonically trapped tagged monomer of a semiflexible polymer under nonequilibrium conditions. %
We performed active Brownian dynamics simulations (Fig.~\ref{fig:simulation}a) in an isothermal Langevin thermostat, where each monomer $i$ experiences an active force $\vv{f}_\te{a}^\perp = \zeta_i v_\perp^{(i)} \hat{\vv{e}}_i(t)$ perpendicular to the polymer's straight ground state and rotational Brownian motion. The decorrelation time $\tau_\te{R}$ was chosen such that the persistence length $v_\perp\tau_\te{R}$ of the active dynamics was significantly smaller than the trap radius $v_\perp/k$, in order to prevent tracer accumulation at the outer margin of the trap of strength $k$~\cite{solon2015abp}. %
To model a situation with spatially heterogeneous activity (Fig.~\ref{fig:simulation}c), we imposed two types of Gaussian-shaped external gradients onto the swim speeds $v_\perp^{(i)}$ of the monomers, corresponding to areas of high liveliness (activity source) or lifeless or dead areas (activity sink) at the position of the central monomer, which serves as the tracer, respectively. %
Complementarily, we performed all-atom molecular-dynamics simulations using the LAMMPS package~\cite{lammps} to simulate the polymer in a nonisothermal Lennard-Jones fluid, as sketched in Fig.~\ref{fig:simulation}b (for details see SI). %
A triangular-shaped temperature profile (red line in Fig.~\ref{fig:simulation}d) was established by maintaining the fluid in designated heating and cooling regions at an elevated ($T_\te{h}$) and lowered temperature ($T_\te{c}$), respectively, using a rescaling algorithm. %
In both approaches, the monomers interacted via WCA and FENE potentials, which renders the chain anharmonic. %

The polymer transmits influences from the heterogeneous (thermal/active) nonequilibrium conditions along its backbone to the tracer at position $\vv{x}(t)$ via an effective driving force $\vv{\Xi}(t)$. %
Similarly, heterogeneous solvent friction along the polymer contour affects the tracer dynamics via a friction kernel $K(t)$
\begin{align}
	\label{eq:tracer-eom}
	\int_{t_0}^t\df t'\, K(t-t') \dot{\vv{x}}(t') = -k \vv{x}(t) + \vv{\Xi}(t)\,,
\end{align}
where $k$ is the trap stiffness. %
While in equilibrium, equipartition among the polymer's normal modes requires that $k\ev{\vv{x}^2} = 2\kBT$, we find the product of trap stiffness and positional tracer fluctuations to be no longer constant, here. At given $k$, an effective temperature $k\ev{\vv{x}^2} = 2\kBT^x$ can be defined by mapping the complex nonequilibrium overdamped dynamics of the tracer particle with its non-Markovian Gaussian noise onto an equivalent equilibrium system at temperature $T^x$ (symbols in Fig.~\ref{fig:simulation}e, f). In the steady-state, the positional probability distribution is given by the generalised Boltzmann factor $P(x) \propto e^{-V/( \kBT^x)}$ (inset of Figs.~\ref{fig:simulation}e, f). %
The overdamped tracer dynamics relaxes with rate $\omega = k^2/(4f\zeta_\perp)$, thus establishing a relation between frequency and trap stiffness (see SI), where $\zeta_\perp$ is the transverse friction coefficient per unit length of the polymer. %

The high-frequency tracer dynamics is governed by high-frequency polymer modes effectively thermalised at the molecular or effective active temperature~\cite{szamel2014self, solon2015abp} in the close vicinity of the tracer (inset of Fig.~\ref{fig:simulation}c, d). %
Specifically, we find $\kBT^x \approx \kBT + \zeta_\perp\tau_\te{R}v_\perp^2(0)/2 = 0.80 \varepsilon$ and $0.98 \varepsilon$, for the activity sink and source, respectively, and $\kBT^x \approx \kBT_\te{h} = 0.9 \varepsilon$ in the case of the nonisothermal bath. %
At lower frequencies, the tracer dynamics is also influenced by longer-wavelength modes, which are more susceptible to the nonuniform heating/activity in the environment. These modes communicate the effective noise contributions of effectively ``cooler'' or ``hotter'' regions far away from the tracer. For the tracer residing in the activity source or in the heating area, the effective temperature of the tracer decreases with decreasing frequency, until it reaches a cut-off corresponding to the longest possible wavelength $\simeq L$ (polymer length). %
Similarly, in the activity sink, the effective temperature will increase with decreasing frequency. %

When the polymer is subject to a linear temperature profile, we obtain an approximately constant effective temperature $\kBT^x$ irrespective of the trap stiffness, as shown by the blue symbols in Fig.~\ref{fig:simulation}f. 
In close analogy to the hydrodynamic description of a Brownian particle in a linear temperature profile~\cite{falasco2016nonisothermal}, the particular symmetry of this setup causes the contributions from the two polymer halves to cancel each other such that the effective temperature is determined by the local thermal energy at the tracer position (SI). %

\emph{Nonequilibrium modal analysis.}\textemdash %
We develop a fully analytical model capable of predicting the observed multiscale active and nonisothermal tracer dynamics (lines in Fig.~\ref{fig:simulation}e and f), within a generalised Langevin framework. %
Starting point of our analysis is the equation of motion for a weakly bending rod with stiffness $\kappa$ and constant backbone tension $f$~\cite{hallatschek2007tension} subject to a nonisothermal temperature profile $T(s)$ and activity profile corresponding to an active force $\zeta_\perp\vv{v}_\perp$. %
The linearised dynamics around the straight ground state in terms of the perpendicular excursions $\vv{r}_\perp(s,t) = (r_\perp^{(1)}(s,t), r_\perp^{(2)}(s,t))$ is given by (for details see SI)
\begin{subequations}
	\begin{align}
		\label{eq:wbr-eom}
		\zeta_\perp \dot{\vv{r}}_\perp (s,t) &= -\kappa \vv{r}_\perp''''(s,t) + f \vv{r}_\perp''(s,t)\,, \nonumber\\
		&\hspace{0.5cm}+ \zeta_\perp \vv{v}_\perp(s,t) + \vv{\xi}_\perp(s,t)\\
		\label{eq:noise-variance}
		\ev{\xi_\perp^{(i)}(s,t) \xi_\perp^{(j)}(s',t')} &= 2 \zeta_\perp \kB T(s) \delta_{ij} \nonumber\\
		&\hspace{0.5cm}\times\delta(s-s') \delta(t-t')\,.
	\end{align}
\end{subequations}
We expand the equation of motion in terms of the polymer normal modes $W_n(s)$ and mode amplitudes $\vv{a}_n(t)$, \ie, $\vv{r}_\perp(s,t) = \sum_n W_n(s) \vv{a}_n(t)$, $ \vv{a}_n(t) = \int\df s\, W_n(s) \vv{r}_\perp(s,t)$, $\vv{v}_n(t) = \int\df s\, W_n(s) \vv{v}_\perp(s,t)$, and $\vv{\xi}_n(t) = \int\df s\, W_n(s) \vv{\xi}_\perp(s,t)$, and obtain
\begin{subequations}
	\begin{align}
		\dot{\vv{a}}_n(t) &= -\frac{\omega_n^2}{\zeta_\perp} \vv{a}_n(t) + \vv{v}_n(t) + \frac{1}{\zeta_\perp} \vv{\xi}_n(t)\,,\\
		\ev{\vv{\xi}_n(t) \cdot \vv{\xi}_m(t')} &= 4 \zeta_\perp \kB \mathcal{T}_{nm}  \delta(t-t')\,,\\
		\ev{\vv{v}_n(t) \cdot \vv{v}_m(t')} &= \mathcal{V}_{nm} e^{ -|t-t'|/\tau_\te{R} }\,.
	\end{align}
\end{subequations}
Here we have defined the thermal and active ``overlap'' $\mathcal{T}_{nm} \equiv \int \df s \, W_n(s) W_m(s) T(s)$ and $\mathcal{V}_{nm} \equiv \int \df s\, W_n(s) W_m(s) v_\perp^2(s) l_0$, respectively, indicating the nonequilibrium-induced crosstalk of the natively orthogonal normal modes. %

\begin{figure}[t]
	\centering
	\includegraphics[width=\columnwidth]{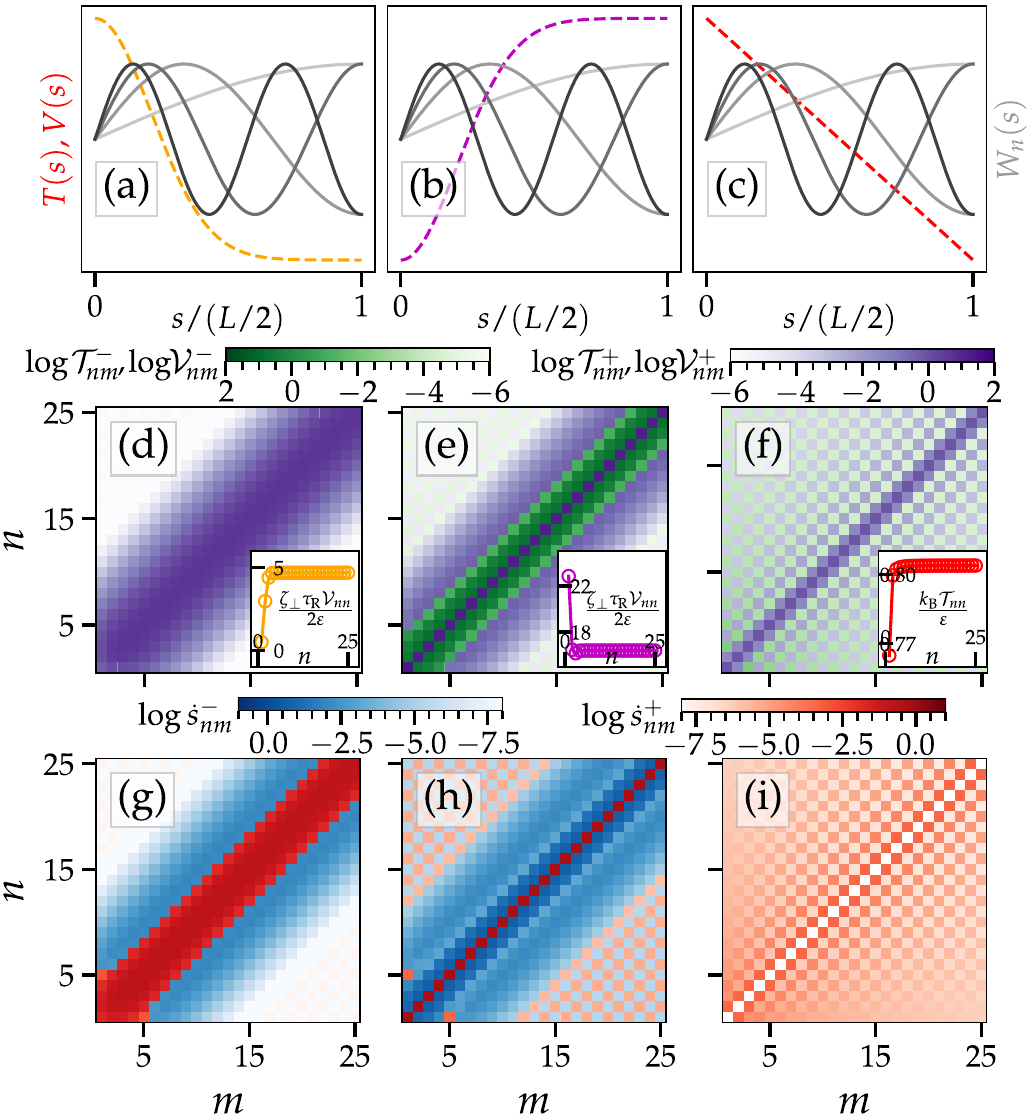}
	\caption{%
		Effective mode temperatures and entropy production rate. %
		(a)-(c)~Mode shape $W_n(s)$ for hinged and torqued ends. Dashed lines depict the activity profiles $v_0(s)$ (a) and (b) or temperature $T(s)$ (c). %
		(d)-(f)~The effective mode temperature matrices $\mathcal{V}_{nm}$ (d)-(e) and $\mathcal{T}_{nm}$ (f) betray the mode-mode coupling and its energy transfer in their off-diagonal components; insets show diagonal values $\mathcal{V}_{nn}$ and $\mathcal{T}_{nn}$. %
		(g)-(i)~The accompanying entropy production rate $\dot{s}_{nm}$ per mode pair $(n,m)$ provides a reference for the multiscale character of the activity. %
	}
	\label{fig:entropy}
\end{figure}

\emph{Intermodal energy transfer and entropy production.}\textemdash %
The steady-state mean potential energy per mode in the absence of activity ($\vv{v}_\perp(s)=0$) is $\omega_n^2\ev{a_n^2} = \kB \mathcal{T}_{nn}$ (see SI) and the quantity $\mathcal{T}_{nn}$ is interpreted as the effective mode temperature, \ie, it encodes how thermal energy is distributed among different polymer modes~\cite{falasco2015energy, yan2017energy}. %
Similarly, the quantity $\zeta_\perp\tau_\te{R}\mathcal{V}_{nm}/2$ can be interpreted as the effective mode temperature due to activity. %
The definition of $\mathcal{T}_{nm}$ and $\mathcal{V}_{nm}$ shows that the mode shape $W_n(s)$ of allowed undulation modes, dictated by the mechanical properties in Eq.~\eqref{eq:wbr-eom} and the corresponding boundary conditions, plays an important role for the mode temperatures. Modes that do not exhibit undulations at positions $s$ cannot tap the corresponding energy supply encoded in $T(s)$ and $\vv{v}_\perp(s)$, as shown in Fig.~\ref{fig:entropy}a-c.
On the other hand, off-diagonal elements determine whether energy stemming from the nonuniform temperature or activity profile can be transferred between the respective normal modes. %
The antisymmetrised correlation function $\ev{\mathcal{L}_{nm}(t)} \equiv \ev{a_n(t_0) a_m(t_0 + t)} - \ev{a_n(t_0) a_m(t_0 - t)}$, sometimes referred to as the angular momentum, can be regarded as a measure of the breaking of detailed balance and is associated with a probability flux in mode-pair space~\cite{gladrow2016broken, gladrow2017nonequilibrium, bacanu2023inferring}. We find that $\ev{\mathcal{L}_{nm}(t)}$ is proportional to the thermal or active overlap $\mathcal{T}_{nm}$ or $\mathcal{V}_{nm}$ for the respective contributions (SI). Thus, for a diagonal system with $\mathcal{T}_{nm} = T_n \delta_{nm}$ and $\mathcal{V}_{nm} = v_n^2 \delta_{nm}$, such as when the polymer is subject to a linear temperature profile, each mode resides in an individual effective equilibrium without intermodal transfer, hence obeying a restricted type of local detailed balance, that could be called ``epipartition''. %

The permanent injection and dissipation of energy at different length scales leads to nonequilibrium fluxes between the polymer modes. %
The off-diagonal effective mode temperatures $\mathcal{T}_{nm}$ and $\mathcal{V}_{nm}$ signal a steady-state energy transfer between polymer modes $n$ and $m$, breaking detailed balance. %
Entropy is produced at various length scales and associated with energy sinks where the injected energy is eventually converted into molecular heat. The entropy production is thus the key-signature of the dissipative energy transfer that accompanies any heterogeneous multiscale activity. %
It is quantified by the entropy production rate $\dot{s}_{nm}$ per mode pair and the total activity by $\dot{S} = \lim_{t\to\infty} \ev{ \ln \mathcal{P}/\mathcal{P}^{(\te{r})}} /t  = \sum_{nm} \dot{s}_{nm}$~\cite{dabelow2019irreversibility, seifert2012stochastic}. %
Here, $\mathcal{P}$ and $\mathcal{P}^{(\te{r})}$ denote the noise path weights and its time-reversed image in mode space, respectively. %
We interpret the active forces $\zeta_\perp \vv{v}_\perp$ as externally applied, \eg, by motor proteins, such that $\vv{v}_\perp$ is even under time-reversal. Such interpretation is appropriate when hydrodynamic effects are neglected and we primarily focus on the network dynamics. %

We find, in general, three contributions to the entropy production rate $\dot{s}_{nm}^{(1,2,3)}$ (see SI). %
On the one hand, the transfer of energy between the modes leads to dissipation at mode $m$ of energy which is injected at mode $n$. The associated entropy production per mode pair $(n,m)$ is proportional to the thermal or active overlap and the squared difference in relaxation timescales $(\tau_m - \tau_n)^2$, ensuring that only cross-terms can contribute:
\begin{subequations}
	\begin{align}
		\label{eq:epr1}
		\dot{s}_{nm}^{(1)} = - &\mathcal{T}_{nm}^{-1} \mathcal{T}_{nm} \frac{(\tau_m - \tau_n)^2}{\tau_m \tau_n (\tau_m + \tau_n)}\,,\\
		\dot{s}_{nm}^{(2)} = - &\mathcal{V}_{nm}^{-1} \mathcal{V}_{nm}\frac{(\tau_m - \tau_n)^2}{(\tau_m + \tau_n) (\tau_m + \tau_\te{R}) (\tau_n + \tau_\te{R})}\,.
	\end{align}
\end{subequations}
Here, $\mathcal{T}_{nm}^{-1}$ and $\mathcal{V}_{nm}^{-1}$ denote the $(n,m)$-th element of the inverse matrix. %
The third contribution is solely found in systems featuring active forces and is the overdamped analogue of entropons~\cite{caprini2023entropons, caprini2023entropy}
\begin{align}\tag{4c}
	\label{eq:epr2}
	\dot{s}_{nm}^{(3)} =  \mathcal{V}_{nm}^{-1} \mathcal{V}_{nm} \frac{1}{\tau_\te{R}}\pt{ \frac{1}{1 + \tau_\te{R}/\tau_m} + \frac{1}{1 + \tau_\te{R}/\tau_n}}\,.
\end{align}
These active elastic elementary excitations, responsible for entropy production, emerge when the active mode-force $\zeta_\perp v_n$ deforms the contour in a way that it encounters considerable friction $\zeta_\perp \dot{a}_n$. Visually, this means that the active force significantly produces excursions from the straight ground state. %
While mode coupling again leads to cross-mode entropy production, this is the only instance where individual modes contribute to $\dot{S}$. %

\emph{Nonequilibrium generalised Langevin equation for a tracer.}\textemdash %
We proceed to calculate the memory kernels $K(t)$ and $\vv{\Xi}(t)$ in Eq.~\eqref{eq:tracer-eom} which sum up how the multiscale energy transfer and dissipation influences the tracer dynamics. %
Technically, we extend the discussion in Ref.~\cite{bullerjahn2011monomer}, assuming that the tracer at position $\vv{x}(t)$ is rigidly tied to the polymer backbone at position $s_0$. A force $\vv{\gamma}(t)$ added to the right-hand side of Eq.~\eqref{eq:wbr-eom} guarantees that $\vv{r}_\perp(s_0,t) \equiv \vv{x}(t)$. %
In order to proceed analytically, we assume that the position of the tracer particle is exactly in the middle of the polymer contour at $s_0=0$ and the polymer is subject to symmetric boundary conditions at its ends at $s=\pm L/2$. %
For details of the derivation see SI. %
Following Ref.~\cite{bullerjahn2011monomer}, we can express the constraint force $\vv{\gamma}(t) = 2\kappa\sum_n \vv{a}_n(t) W_n'''(0^+)$ in terms of the polymer modes by insisting on the continuity of the contour at position $s_0$. Here, $0^+$ denotes the limit $s \to 0$ from $s > 0$. %
Since the constraint force $\gamma(t)$ is the force exerted by the polymeric degrees of freedom onto the tracer and \emph{vice versa}, it can be written in terms of the polymer contributions in Eq.~\eqref{eq:tracer-eom}, \ie, $\vv{\gamma}(t)=-\vv{\Xi}(t)+\int_{t_0}^t\mathrm{d}t'\,K(t-t')\dot{\vv{x}}(t')$. %
After solving the tracer-polymer compound mode amplitude equation for $\vv{a}_n(t)$ and the corresponding eigenvalue problem for $W_n(s)$, $\gamma(t)$ and thus $K(t)$ and $\vv{\Xi}(t)$ can be directly inferred (see SI). %

If we neglect their effect on the solvent friction, the nonisothermal temperature and activity profiles solely enter the noise term $\vv{\Xi}(t)$, and the deterministic memory kernel $K(t)$ remains identical to the equilibrium case~\cite{bullerjahn2011monomer}. %
The strength of the nonequilibrium coloured Gaussian noise $\vv{\Xi}(t)$ is composed of a convolution of contributions from all modes such that we continue the analysis in Fourier space for $\vv{\Xi}(t)$ to decipher the individual contributions. Then, $K_\omega = \int_{-\infty}^\infty\df t\, K(t) e^{-i\omega t}$ is the Fourier transform of the memory kernel and $\vv{\Xi}_\omega$ is the Fourier transform of the noise term $\vv{\Xi}(t)$. %
The zero-mean ($\ev{\vv{\Xi}_\omega} = 0$) Fourier noise can be characterised by $\ev{ \Xi_\omega^{(i)} \Xi_{-\omega'}^{(j)} } = K_\omega k_\mathrm{B} \mathscr{T}_\omega \delta_{ij}\delta(\omega -\omega')$, with $i$ and $j$ the two directions of deflection. %
Here we have defined the frequency-dependent spectral noise temperature $\mathscr{T}_\omega$. It encodes how strongly the nonisothermal driving and the active forces couple to the tracer dynamics at frequency $\omega$. It explicitly depends on the mode relaxation times $\tau_n = \omega_n^2/\zeta_\perp$ and their eigenfunctions $W_n(s)$ and thus on the boundary conditions at $s = \pm L/2$. %
However, for $L \to \infty$ or, equivalently, $t \ll \tau_1$ ($\omega \gg \omega_1$), perturbations from the boundaries become negligible and the tracer dynamics is dictated by the internal degrees of freedom of the polymer. %
This admits an explicit analytical universal expression for $\mathscr{T}_\omega$ independent of the boundary conditions~\cite{hallatschek2005propagation} that becomes exact in the respective limit (SI). %
In this way, our formalism can be readily generalised to diverse viscoelastic systems subject to activity or nonisothermal temperature profiles, as encountered in the cytoplasm of living cells~\cite{guo2014probing} or active motor-filament networks~\cite{gladrow2016broken}, without the need to account for potentially intricate (unknown) boundary conditions. %
The noise spectrum consists of contributions from all modes via the overlap $\mathcal{T}_{nm}$ and $\mathcal{V}_{nm}$
\begin{align}
	\label{eq:effective_noise_temperature_dissipation}
	\mathscr{T}_\omega = \frac{\sum\limits_{n,m} \phi_{nm}(\omega) \mathcal{T}_{nm} }{\sum_n \phi_{nn}(\omega)}
	+ \frac{\sum\limits_{n,m} \phi_{nm}(\omega)\frac{1}{1+\tau_\mathrm{R}^2\omega^2} \frac{\zeta_\perp\tau_\mathrm{R}\mathcal{V}_{nm}}{2}}{\sum_n \phi_{nn}(\omega)}\,,
\end{align}
weighted by the dissipation function $\phi_{nm}(\omega)$ (SI).
Its diagonal elements $\phi_{nn}(\omega)$ represent the contribution of the $n$-th polymer mode to the memory kernel $K_\omega = \sum_n \phi_{nn}(\omega)$. %
The two contributions in Eq.~\eqref{eq:effective_noise_temperature_dissipation} stem from the nonisothermal temperature and activity profiles and can be treated separately. %
The rotational decorrelation timescale $\tau_\te{R}$ of the active force acts as a frequency cutoff, such that, at frequencies $\omega \gg 1/\tau_\mathrm{R}$, the active forces barely change the configuration of the polymer and thus do not contribute to the noise spectrum, in agreement with experimental findings~\cite{fakhri2014high}. %
Notably, both the effective noise spectrum $\mathscr{T}_\omega$ in Eq.~\eqref{eq:effective_noise_temperature_dissipation} and the entropy production rates in Eqs.~\eqref{eq:epr1}-\eqref{eq:epr2} depend on the overlap $\mathcal{T}_{nm}$ and $\mathcal{V}_{nm}$, revealing that the tracer's (active) noise spectrum is a quantitative manifestation of broken detailed balance and driver of the entropy production. %

\emph{Sensing multiscale activity in tracer fluctuations.}\textemdash %
Equipped with explicit expressions for $\vv{\Xi}$ and $K$ Eq.~\eqref{eq:tracer-eom} allows us to analytically calculate the tracer fluctuations $\ev{\vv{x}^2} = 2\kBT^x/k$. Using the Wiener-Khintchin theorem~\cite{chaikin1995principles}, we find $\kBT^x = \frac{k}{2\pi} \int_{-\infty}^{\infty}\mathrm{d}\omega\, \mathscr{T}_\omega |R_\omega^x|^2 K_\omega$, where $R_\omega^x = 1 /(i \omega K_\omega^+ + k)$ is the system's response function, and $K_\omega^+$ the half Fourier transform $K_\omega^+ = \int_{0}^\infty\df t\, K(t) e^{-i\omega t}$. 
The overdamped oscillator is detuned with respect to the standard underdamped resonance frequency $\sqrt{k/m}$ and, since equipartition is broken, the different frequencies in the noise spectrum contribute unequally to the dynamics. The memory kernel and the harmonic potential selectively filter the noise spectrum. %
We employ the inverse relaxation timescale $\omega = k^2/(4f\zeta_\perp)$ in order to estimate $T^x = \frac{k \ev{\vv{x}^2}}{2\kB} \simeq \mathscr{T}_{\omega(k)}$, such that we can infer $\mathscr{T}_{\omega}$. %
This transforms the harmonic trap into a spectroscopic tool to scan the system's nonequilibrium modes by varying the trap stiffness $k$. %
The comparison of simulation results for $\kBT^x$ (symbols) and the predicted spectrum $\mathscr{T}_{\omega}$ (lines) in Fig.~\ref{fig:simulation}e and f proves the applicability of the method to tackle the inverse problem of inferring characteristics of multiscale activity in a complex nonequilibrium environment. %
Particularly, we have exactly shown that activity at different scales couples to the dynamics at different frequencies, such that an image of the activity landscape can be obtained by scanning through frequencies. %
Our work thereby provides a proof of principle for such nonequilibrium thermospectroscopy~\cite{geiss2020thermometry, falasco2016nonisothermal} in polymeric systems, where frequency scales become experimentally accessible through tunable parameters like backbone stiffness, solvent friction~\cite{mollenkopf2023heavy}, and imposed tension. %
Moreover, the general framework could be directly generalised to various active-matter systems, such as flickering cell membranes, the (in-)animate character of which have been a subject of century-long debates~\cite{turlier2016equilibrium}. %

\emph{Summary.}\textemdash %
In this Letter, we have laid the theoretical foundation of a nonequilibrium thermospectroscopic method applicable to active matter and nonequilibrium colloidal systems with multiscale activity. %
Our work establishes a conceptual link between the analysis of the breaking of detailed balance, mode-energy transfer, entropy production, and the effective temperatures in scale-dependent nonequilibrium or active biological and synthetic networks.
We have explicitly proven the applicability of the spectroscopic tool for the analysis of scale-dependent activity, providing a conceptual paradigm of how to decipher scale-dependent biological functions and their impact on, \eg, physiology in both healthy and diseased states~\cite{guo2014probing}. %

\section{Acknowledgements}
We thank Sven Auschra, Jakob T\'omas Bullerjahn, and Gianmaria Falasco for valuable discussions in the early stages of this project.



%
\end{document}